\documentclass[12pt]{iopart}
\usepackage{amssymb}
\usepackage{iopams,bm,verbatim}
\usepackage[T1]{fontenc}
\def\Journal#1#2#3#4{{#4} {\it #1} {\bf #2}, #3 }
\def\Psic{\overline{\Psi}_2}
\def\Nc{\overline{N}}
\def\Xc{\overline{X}}

\newcommand{\w}[1]{\bm{#1}} 
\newcommand{\vrk}{\hfill $\Box$}
\def\tho{\textrm{\TH}}

\begin{document}

\title{Petrov type $D$ pure radiation fields \\ of Kundt's class}

\author{L.~De Groote, N.~Van den Bergh and L.~Wylleman}

\address{Ghent University, Department of Mathematical Analysis IW16, \\ Galglaan 2, 9000 Ghent, Belgium}

\begin{abstract}
    We present all Petrov type $D$ pure radiation space-times, with or without cosmological constant, with a shear-free, non-diverging geodesic principal null congruence.
\end{abstract}

\section{Introduction}
A space-time is said to belong to Kundt's class if it admits a null congruence, generated by a vectorfield $\tilde{k}$, which is non-diverging (i.e.~which is both non-expanding and non-rotating and therefore also geodesic). In the case of a pure radiation field, which, by definition, has an energy-momentum tensor of the form $T_{ab} = \phi k_a k_b,\; \mathrm{with} \; k_ak^a =0,\,\phi\neq 0$, one can furthermore show\cite{Kramer}, making use of the null energy condition $R_{ab}k^a k^b \geq 0$ and the Raychaudhuri-equation
\begin{equation*}
    \Theta_{,a}\tilde{k}^a - \omega^2 + \Theta^2 + \sigma\overline{\sigma} = -\frac{1}{2}R_{ab} \tilde{k}^a \tilde{k}^b,
\end{equation*}
that $k$ and $\tilde{k}$ are aligned and that the associated null congruence is shear-free :
\begin{eqnarray*}
    R_{ab}\tilde{k}^a \tilde{k}^b = - 2 \sigma \overline{\sigma} \qquad &\Rightarrow \sigma = 0
    \quad \mathrm{and} \quad R_{ab} \tilde{k}^a \tilde{k}^b = 0, \\
    R_{ab} = \phi k_a k_b +\Lambda g_{ab}&\Rightarrow k \propto \tilde{k},
\end{eqnarray*}
where $\Lambda$ is the cosmological constant. The Goldberg-Sachs theorem implies then that the Kundt space-times must be algebraically special (type $II$, $D$, $III$ or $N$ or conformally flat) and that $\tilde{k}$ is aligned with a repeated principal null direction of the Weyl tensor.

Conversely, for aligned \emph{Petrov type $D$} pure radiation fields, it follows immediately from the Bianchi equations (without invoking the energy conditions) and from
\begin{eqnarray}
    \label{psis}
    \Psi_0 = \Psi_1 = \Psi_3 = \Psi_4 = 0, \qquad &\Psi_2 \neq 0,\\
    \label{phis}
    \Phi_{00}=\Phi_{11}=\Phi_{01}=\Phi_{12}=\Phi_{02}=0, \quad &\Phi_{22}\neq 0,
\end{eqnarray}
that
\begin{equation}
    \label{kappa}
    \kappa = \sigma = \lambda = 0,
\end{equation}
such that the principal null direction $k$ aligned with the radiative direction is geodesic and shear-free, i.e.\, the converse part of the Goldberg-Sachs is generalized to this situation. Moreover, it is known that aligned Petrov type $D$ pure radiation fields cannot be of (null) Maxwell type\cite{DebeverVdBLeRoy} nor of (null) neutrino or scalar field type\cite{WilsVdB}, while Wils\cite{Wils} showed that $k$ is necessarily non-twisting ($\overline{\rho}=\rho$). The diverging solutions ($\rho\neq 0$) then belong to the Robinson-Trautman class and are all explicitly known\cite{Kramer}. In the non-diverging case $\rho=0$, however, only a few type $D$ examples are available\cite{WilsVdB}; here we present all solutions belonging to the latter class.
\\$ $\\
As shown by Kundt\cite{Kundt} in 1961 the line-elements admitting a geodesic, shear-free and non-diverging null congruence can all be expressed in the form:
\begin{equation}
	\label{dskundt}
	\mathrm{d} s^2 = 2 P ^{-2} \mathrm{d}\zeta \mathrm{d}\overline{\zeta}- 2 \mathrm{d}u (\mathrm{d}r + H\mathrm{d}u + W\mathrm{d}\zeta + \overline{W} \mathrm{d} \overline{\zeta}),
\end{equation}
in which $P$ is a real function of $(\zeta, \overline{\zeta},u)$ and $H, W$ are respectively real and complex functions of $(\zeta,\overline{\zeta},u,r)$, to be determined by appropriate field equations. The vacuum solutions of this type have been known for a very long time\cite{Kinnersley}. A procedure (based on Theorem 31.1 in~\cite{Kramer}) is also available allowing one to generate non-vacuum solutions of Kundt's class, from vacuum solutions. However, in this way the Petrov type of the metric generally will be changed from $D$ to $II$. Insisting that the Petrov type does not change, constrains the function $H_0$ of the `background-metric' and it was not certain whether one could generate all pure radiation solutions in this way. This technique was used in~\cite{WilsVdB}, where the authors managed to construct a family of type $D$ pure radiation fields and where they conjectured that these solutions were the only aligned type $D$ pure radiation fields of Kundt's class. Below we show that the solutions obtained in~\cite{WilsVdB} only cover a small part of the entire family, which we present in this paper.

The most obvious way to find all type $D$ pure radiation fields of Kundt's class, is to start from the general Kundt metric (\ref{dskundt}), express that the solutions we are looking for are of Petrov type $D$, and to make sure that the field equations for pure radiation are satisfied. This however introduces a hard to solve system of non-linear conditions. We therefore prefer to start from the general Geroch Held Penrose (GHP)\cite{GHP} or Newman Penrose (NP)\cite{NP} equations, where we follow the notation and conventions of~\cite{Kramer}, and to extract from these all possible invariant information, before introducing any coordinates.

\section{GHP analysis}

Introducing the set of conditions (\ref{psis})-(\ref{kappa}) and the non-diverging condition $\rho=0$ within the GHP Bianchi equations, one arrives at \\
\begin{minipage}{0.4\textwidth}
    \begin{eqnarray}
        \nonumber\tho\Psi_2 = 0,\\
        \nonumber\eth \Psi_2 = 3 \tau \Psi_2,\\
        \nonumber \eth '\Psi_2 = - 3 \pi \Psi_2,\\
        \nonumber \tho ' \Psi_2 = -3\mu\Psi_2,
    \end{eqnarray}
\end{minipage}
\begin{minipage}{0.59\textwidth}
    \begin{eqnarray}
        \nonumber\tho \Phi_{22} = 0,\\
        \label{ethPhi22}\eth \Phi_{22} = 3 \overline{\nu}\Psic + \tau\Phi_{22},\\
        \nonumber\eth '\Phi_{22} = 3\nu\Psi_2 + \overline{\tau} \Phi_{22}.\\
        \nonumber
    \end{eqnarray}
\end{minipage}\\$ $\\
The GHP Ricci equations then reduce to\\
\begin{minipage}{0.4\textwidth}
    \begin{eqnarray}
        \tho \tau = 0,\nonumber\\
        \nonumber \eth \tau = \tau^2,\\
        \nonumber \eth ' \tau = \tau \overline{\tau}+\Psi_2 + \frac{\Lambda}{3},
    \end{eqnarray}
\end{minipage}
\begin{minipage}{0.59\textwidth}
    \begin{eqnarray}
        \nonumber\eth ' \mu = (\overline{\mu} - \mu) \pi,\\
        \label{ethnu}\eth ' \nu = (\overline{\tau} - \pi)\nu,\\
        \nonumber \eth ' \pi = -\pi^2,
    \end{eqnarray}
\end{minipage}\\
and
\begin{eqnarray}
     \nonumber \tho \nu = \tho ' \pi + (\overline{\tau} + \pi)\mu,\\
     \label{Deltamu} \tho ' \mu = \eth \nu - \mu^2 + \overline{\nu}\pi -\nu \tau -\Phi_{22},\\
     \nonumber \tho \mu = \eth \pi + \pi \overline{\pi} + \Psi_2 +\frac{\Lambda}{3}.
\end{eqnarray}
The remaining basic variables at this stage are $\mu, \nu, \pi, \tau, \Psi_2$ and their complex conjugates, together with $\Phi_{22}$; the cosmological constant $\Lambda$ appears as a parameter. From the integrability conditions for $\Psi_2$ one obtains : \\
\begin{minipage}{0.4\textwidth}
    \begin{eqnarray*}
        \tho \nu = 0,\\
        \tho \mu = \pi \overline{\pi} - \tau \overline{\tau},\\
        \eth \mu = -\tho ' \tau,\\
    \end{eqnarray*}
\end{minipage}
\begin{minipage}{0.59\textwidth}
    \begin{eqnarray}
        \nonumber \tho \pi = 0,\\
        \label{deltapi}\eth \pi = -(\tau \overline{\tau} +\Psi_2 +\frac{\Lambda}{3}),\\
        \nonumber\tho ' \pi = -(\overline{\tau} + \pi)\mu.\\\nonumber
    \end{eqnarray}
\end{minipage}\\
$[\eth ', \tho]\,\mu$ now yields the algebraic equation
\begin{equation}
\label{vgl1}
    \left( 2\tau\overline{\tau} + \Psic + \frac{\Lambda}{3}\right) \pi
    + \left( 2\tau\overline{\tau} + \Psi_2 + \frac{\Lambda}{3}\right) \overline{\tau}=0.
\end{equation}
From $[\eth,\eth ']\mu$ we can extract an expression for $\eth '\tho '\tau$ :
\begin{equation*}
    \eth ' \tho ' \tau=\left(\Psic-2\Psi_2 - 2 \overline{\tau}\tau -\frac{\Lambda}{3} \right) \mu
    + \left(\Psi_2 + 2 \overline{\tau}\tau + \frac{\Lambda}{3} \right) \overline{\mu} - \pi \tho ' \tau.
\end{equation*}
If one then calculates $[\eth, \eth ']\,\tho '\tau$, making use of $[\tho ', \tho]\tau$, $[\eth, \tho ']\tau$, and (\ref{vgl1}) the following expression for $\tho ' \tau$ is obtained,
\begin{equation}
    \tho ' \tau =  \overline{\mu} \tau -\mu\overline{\pi} - 2\mu \tau,
\end{equation}
after which $[\tho ', \eth] \,\pi$ results in a second algebraic equation
\begin{eqnarray}
\label{vgl2}
    \mu Q = 0, \qquad Q = 6(\overline{\tau}\overline{\pi}+\tau \pi ) + 3(\tau \overline{\tau} + \pi \overline{\pi} )
    - \Lambda + 3(\Psi_2 + \Psic).
\end{eqnarray}
$ $\\
From $\eth (\ref{vgl1})-2\tau (\ref{vgl1})$ one readily infers
\begin{equation}\label{pitaunormsequal}
    \pi \overline{\pi} = \tau \overline{\tau}.
\end{equation}
In the case where $\pi=\tau=0$, $\eth\pi=0$ and $\tho '\eth\pi=0$ yield
\begin{equation}\label{casepi0}
	\Psi_2=-\frac{\Lambda}{3}, \quad \mu=0
\end{equation}
and all the above equations are identically satisfied. Consider now the remaining case $\pi\tau\neq 0$, and introduce the zero-weighted (i.e.\ spin- and boost-invariant) quantities
\begin{eqnarray}
	A \equiv |\pi\overline{\pi}|^{1/2}=|\tau\overline{\tau}|^{1/2}&&\quad(\overline{A} = A >0),\label{defA}\\
	T\equiv \frac{\pi}{\overline{\tau}}=\frac{\tau}{\overline{\pi}}&&\quad\left(\overline{T} = T^{-1}\right).\label{defT}
\end{eqnarray}
In this way $\pi$ and $\tau$ are conveniently parametrized by
\begin{equation}
	\pi=\frac{A}{B},\quad \tau=ATB,
\end{equation}
the weight of both being absorbed in the (1,-1)-weighted quantity
\begin{eqnarray}
	B\equiv \left(\frac{\overline{\pi}}{\pi}\right)^{1/2}=\left(\frac{\tau}{\overline{\tau}}\right)^{1/2}&&\quad (\overline{B}=B^{-1}).
\end{eqnarray}
Within this notation, the equations (\ref{vgl1}) and (\ref{vgl2}) translate into
\begin{eqnarray}
	(T+1)(2A^2+\Lambda/3)+\Psi_2+T\Psic=0,\label{vgl1b}\\
	\mu Q=0,\qquad Q\equiv 2(T^2+T+1)A^2+T(\Psi_2+\Psic-\Lambda/3),\label{vgl2b}
\end{eqnarray}
respectively. Furthermore, we define the (-2,-2)-weighted quantities
\begin{equation}
	X\equiv \eth\nu,\quad N\equiv B\nu
\end{equation}
and the zero-weighted derivative operators
\begin{equation}
	{\cal D}\equiv B^{-1}\,\eth,\quad {\cal D}'\equiv B\,\eth'.
\end{equation}
From (\ref{ethPhi22})-(\ref{Deltamu}), $[\tho ' , \eth]\mu$ and $[\eth, \eth ']\nu$ one then deduces the following autonomous first-order system for the zero-weighted fields
$\Psi_2,\,\Psic,\,T,\,A$  and the (-2,-2)-weighted fields
$X,\,\overline{X},\,N,\,\overline{N},\,\Phi_{22}$:
\begin{eqnarray}
	{\cal D}\Psi_2=3AT\Psi_2,\quad{\cal D}\Psic=-3A\Psic,\label{DPsi2}\\
	{\cal D}T=AT(T+1), \quad {\cal D}A=-A^2-\frac{3\Psi_2+\Lambda}{6},\label{DTA}
\end{eqnarray}
\begin{eqnarray}
	{\cal D}X&=&A(4T+1)X-AT\overline{X}-2A^2T(T+1)N+2A^2(T+2)\overline{N}A^2\nonumber\\
	&&+(3\Psic+\Psi_2+\Lambda/3)\overline{N}-A(T+1)\Phi_{22},\label{DX}\\
	{\cal D}\overline{X}&=&A(T-1)\overline{X}+(4\Psic+\Lambda/3+2A^2)\overline{N},
\end{eqnarray}
\begin{eqnarray}
	{\cal D}N=X+\frac{(3\Psi_2+\Lambda)}{6A}N,\quad
	{\cal D}\overline{N}=\left(\frac{A(T-1)}{T}-\frac{3\Psic+\Lambda}{6A}\right)N,\\
	{\cal D}\Phi_{22}=AT\Phi_{22}+3\Psic\Nc.\label{DPhi22}
\end{eqnarray}
For later use, one immediately observes from (\ref{DTA}) that \emph{$T$ constant implies $T=-1$}. Also notice that (\ref{DPsi2}),(\ref{DTA}) forms a zero-weighted subsystem, whereas the equations (\ref{DX})-(\ref{DPhi22}) are (-2,-2)-weighted, such that their right hand sides are necessarily linear and homogeneous in $X,\Xc,N,\Nc,\Phi_{22}$. The complex conjugates of (\ref{DPsi2})-(\ref{DPhi22}) yield an analogous ${\cal D}'$ system. However, as deduced from $[\eth,\eth']\Phi_{22}$, both systems are constrained by
\begin{eqnarray}
    \label{la1} 3T(\Psi_2 X - \Psic\Xc)+ 6A(T^2\Psi_2 N -\Psic \Nc)+ T(\Psi_2 - \Psic)\Phi_{22}  = 0,
\end{eqnarray}
which is again linear and homogeneous in $X,\Xc,N,\Nc,\Phi_{22}$.

We are now ready to prove:\\

\textbf{Theorem} \emph{W.r.t.\ a Weyl canonical tetrad (\ref{psis}),(\ref{phis}) any aligned Petrov type $D$ pure radiation field, for which the principal null direction, aligned with the radiative direction is non-diverging, satisfies the following boost- and rotation-invariant properties:}
\begin{eqnarray}
	\kappa=\sigma=\rho=0,\quad \lambda=\mu=0,\quad \nu\neq 0,\label{fff}\\
	\pi+\overline{\tau}=0,\quad \Psic=\Psi_2,\quad
	\pi\,\Phi_{22}+\nu\left(\Psi_2+\frac{\Lambda}{3}\right)=0.\label{ggg}
\end{eqnarray}
\textsc{Proof} It immediately follows from (\ref{Deltamu}) that $\mu=\nu=0$ is inconsistent with $\Phi_{22}\neq 0$, and from (\ref{vgl1}) that $\pi+\overline{\tau}=0$ implies $\Psi_2$ to be real. By (\ref{casepi0}) the theorem is true in the case where $\pi=\tau=0$, and  it remains to establish $\mu=0$ and $\pi=-\overline{\tau}$ ($T=-1$ in the above notation) when $\pi\tau\neq 0$.

Referring to (\ref{vgl2b}), we first treat the case $Q=0$. Two linear and homogeneous systems in $X,\Xc,N,\Nc$ and $\Phi_{22}$ are obtained from [(\ref{la1}), ${\cal D}$(\ref{la1}), ${\cal D}'$(\ref{la1}), ${\cal D}{\cal D}'$(\ref{la1})] and ${\cal D}{\cal D}$(\ref{la1}), respectively ${\cal D}'{\cal D}'$(\ref{la1}). As $\Phi_{22}\neq 0$, the determinants of these systems must vanish. This yields two polynomial relations in $A^2,\,\Lambda,\Psi_2,\,\Psic$ and $T$; eliminating $A^2$ and $\Lambda$ by means of (\ref{vgl1b}) and $Q=0$, and appropriately scaling the results with $\Psi_2$, yields two polynomial relations in the zero-order variables $T$ and $k\equiv\Psic/\Psi_2$ of the form $(T+1)^3T^7P_i(T,k)=0,\,i=1,2$, where $P_1$ and $P_2$ are irreducible and non-proportional.
Thus
$T$ is constant, whence $T=-1$ and $\Psic=\Psi_2$. Substituting this into $Q=0$ yields $\pi\overline{\pi}-\Psi_2+\Lambda/6=0$, and calculating the $\tho'$-derivative hereof gives  $\mu = 0$.

Thus $\mu=0$ in any case, and (\ref{Deltamu}) implies 
\begin{equation}
	\label{extramu}
	\eth'\overline{\nu}=\eth\nu=-\pi\overline{\nu}-\overline{\pi}\nu+\Phi_{22},\quad\textrm{i.e.,}\quad \Xc=X=-A(\Nc-TN)+\Phi_{22}.
\end{equation}
Substituting this in
[(\ref{la1}),${\cal D}$(\ref{la1}),${\cal D}'$(\ref{la1})], [(\ref{la1}),${\cal D}$(\ref{la1}),${\cal D}{\cal D}$(\ref{la1})] and [(\ref{la1}),${\cal D}'$(\ref{la1}),${\cal D}'{\cal D}'$(\ref{la1})] yields three linear and homogeneous systems in $N,\Nc$ and $\Phi_{22}$. Eliminating $A^2$ and $\Lambda$ from the three determinants and (\ref{vgl1b}) leads to two  polynomial relations in $T$ and $k$ with $(T+1)^3$ as the only common factor, such that again $T=-1$ and $\Psic=\Psi_2$. Inserting this and (\ref{extramu}) into (\ref{la1}) implies $\Nc=N$ (i.e., $\nu/\pi$ is real), a final ${\cal D}$-derivative of which yields 
\begin{equation*}
	A\Phi_{22}=N(\Psi_2+\Lambda/3), 
\end{equation*} 
the translation of the last equation of (\ref{ggg}).\vrk\\ 

One checks that, given the specifications (\ref{psis}),(\ref{phis}) and (\ref{fff}),(\ref{ggg}), the remaining GHP equations are consistent,
such that corresponding solutions exist. Referring to (\ref{vgl2b}) and the above proof, we finally remark that all derivatives of $Q\,\Psi_2^{-2/3}$ are zero, such that one has the following important relation:
\begin{equation}
    \label{vgl}
    \pi \overline{\pi} =\Psi_2 -\frac{\Lambda}{6} + c_1 \Psi_2^{2/3}, \qquad c_1 \quad
    \mathrm{constant}.
\end{equation}
Because of (\ref{casepi0}), this only yields new information when $\pi\neq 0$; the case $c_1=0\Leftrightarrow Q=0$ will be distinguished from $c_1\neq0$ for the corresponding integration in $\S$ \ref{subsection: pi nonzero}.

\section{NP analysis}

In order to construct the metrics, we now switch to the NP-formalism, where we first introduce the invariant information, as obtained above. Appropriate boosts and rotations allow one to put $\epsilon = 0$ and $ \beta = -(\overline{\alpha} +\pi )$. This fixes the null tetrad up to rotations and boosts satisfying $\mathrm{D}\theta = 0$ and $\mathrm{D} A = 0 = \delta A = \overline{\delta} A$. As $\nu \neq 0$ and $\mathrm{D}\nu = (\tho -3\epsilon-\overline{\epsilon})\nu= 0$, one can completely eliminate the rotational degree of freedom by requiring $\overline{\nu} = \nu$. By (\ref{ggg}) this implies $\overline{\pi} = \pi$. The solutions can then be divided into two families : one in which $\pi = 0$ and one in which $\pi \neq 0$. We also introduce $\gamma_0$ and $\gamma_1/\nu$, the real and the imaginary part of $\gamma$, respectively.

\subsection{The family $\pi = 0$}

In this case (\ref{deltapi}) shows $\Psi_2 = - \frac{\Lambda}{3} \neq 0$. The NP Ricci equations can then be rewritten in the form :
\begin{equation}
	\label{phiopl}
	\Phi_{22} = -4\alpha\nu,
\end{equation}
\begin{minipage}{0.4\textwidth}
    \begin{eqnarray}
        \nonumber\mathrm{D}\alpha = 0, \\
        \label{Dgamma}\mathrm{D}\gamma_0 = -\frac{\Lambda}{2},\\
        \nonumber \mathrm{D}\gamma_1 = 0,
    \end{eqnarray}
\end{minipage}
\begin{minipage}{0.59\textwidth}
    \begin{eqnarray}
		\nonumber \nonumber\mathrm{D}\nu = 0,\\
        \label{dnu} \overline{\delta}\nu = - 2 \alpha \nu,\\
		\nonumber \delta \nu = -2 \alpha\nu,
    \end{eqnarray}
\end{minipage}
and
\begin{eqnarray}
    \label{rest1} \delta \gamma_0 = 0,\\
    \label{rest2} \nu \Delta \alpha - \mathrm{i} \overline{\delta} \gamma_1 = 0,\\
    \label{rest3} \delta \alpha + \overline{\delta}\alpha = 4 \alpha^2 + \frac{\Lambda}{2}.
\end{eqnarray}
The Bianchi-equations now yield an expression for $\delta \alpha$ (by which (\ref{rest3}) becomes an identity) :
\begin{equation*}
    \delta \alpha = 2 \alpha^2 + \frac{\Lambda}{4}.
\end{equation*}

By (\ref{phiopl}) and (\ref{dnu}) $\nu$ cannot be constant and this allows one to use $\nu$ as a coordinate. As is clear from (\ref{Dgamma}), $\gamma_0$ provides a second independent function and hence can be used to define a coordinate $r=-2\gamma_0/\Lambda$. As $\delta \gamma_0 = 0$ it follows that $\w{\omega}^4 = \mathrm{d}r + H \w{\omega}^3$ ($H = -\Delta r$, $\w{\omega}^i$ representing the basis one-forms). From (\ref{dnu}) we see that we can write $\w{\omega}^1 + \w{\omega}^2$ as :
\begin{equation}
    \label{om1+om2}
    \w{\omega}^1 + \w{\omega}^2 = \frac{1}{2} \frac{\Delta \nu \w{\omega}^3- \mathrm{d} \nu}{\alpha \nu}.
\end{equation}
The Cartan-equations read
\begin{eqnarray}
    \label{dom1} \mathrm{d} \w{\omega}^1 &=& 2 \alpha \, \w{\omega}^1 \wedge \w{\omega}^2 + 2\mathrm{i}\gamma_1/\nu \, \w{\omega}^1 \wedge \w{\omega}^3, \\
    \nonumber \mathrm{d} \w{\omega}^3 &=& 0, \\
    \nonumber \mathrm{d} \w{\omega}^4 &=& \nu (\w{\omega}^1 + \w{\omega}^2)\wedge \w{\omega}^3 - 2\gamma_0 \, \w{\omega}^3 \wedge \w{\omega}^4,
\end{eqnarray}
showing that a function $u$ exists, such that $\w{\omega}^3 = \mathrm{d}u$. A fourth coordinate $x$ is defined by writing $\w{\omega}^1-\w{\omega}^2=\mathrm{i}(N\mathrm{d}\nu + U\mathrm{d}u + 4\nu X\mathrm{d}x)$, with $N, U, X$ real functions of $(u,r,x,\nu)$, where the factor $\nu$ has been introduced for later convenience. From (\ref{dom1}) it then follows that the functions $N, U $ and $X$ are independent of $r$, so a transformation of $x$ can be used to put $N=0$. Hence we obtain the following basis one-forms :
\begin{eqnarray}
    \label{ome1} \w{\omega}^1 &= \frac{2\mathrm{i}\alpha \nu U  + \Delta \nu }{4 \alpha \nu} \mathrm{d}u - \frac{1}{4 \alpha \nu} \mathrm{d} \nu + 2 \mathrm{i} \nu X \mathrm{d}x, \\
    \nonumber \w{\omega}^3 &= \mathrm{d}u , \\
    \nonumber \w{\omega}^4 &= \mathrm{d}r + H \mathrm{d}u,
\end{eqnarray}
where $U$ and $X$ are real functions of $(u, \nu, x)$ and $H$ is a real function of $(u, r, \nu, x)$. The corresponding directional derivatives are then given by :
\begin{eqnarray*}
    \delta = -2\alpha\nu \frac{\partial}{\partial \nu} - \frac{\mathrm{i}}{4\nu X} \frac{\partial}{\partial x}, \\
    \Delta = \frac{\partial}{\partial u} - H \frac{\partial}{\partial r} + \Delta\nu\frac{\partial}{\partial \nu} - \frac{U}{4\nu X} \frac{\partial}{\partial x}, \\
    \mathrm{D} = \frac{\partial}{\partial r}.
\end{eqnarray*}
Note that, because of the remaining boost freedom, $r$ is fixed only upto transformations of the form $r \rightarrow g_1(u)r+g_2(u)$. 

From the directional derivatives of $\alpha$ we see that $\alpha$ can be written in the form
\begin{equation}
	\label{alphaopl}
	\alpha = \frac{\sqrt{-2\Lambda\nu^2 + s m(u)^2}}{4 \nu}, \qquad \quad s=\pm 1.
\end{equation} 
Applying $[\Delta, \mathrm{D}]$ and $[\delta, \Delta]$ to $r$, we obtain an expression for $H(u,r,\nu,x)$ :
\begin{equation*}
    H(u,r,\nu,x) = -\frac{\Lambda r^2}{2} +\frac{4 \alpha \nu}{\Lambda} + n(u).
\end{equation*}
From $[\overline{\delta}, \delta]x$ we get $X(u,\nu,x) = X(u,x)$, while $[\delta,\Delta]x$ shows that
\begin{equation}
    \label{Cu}
    \frac{\partial X}{\partial u} = \frac{1}{4\nu} \left(\frac{\partial U}{\partial x}- 4 X \Delta \nu \right)
\end{equation}
and
\begin{equation}
    \label{Bnu}
    \frac{\partial U}{\partial \nu} = \frac{U}{\nu} - \frac{\gamma_1}{\alpha\nu^2}.
\end{equation}
From (\ref{rest2}) and $\mathrm{D}\gamma_1$, we see that $\gamma_1 = \gamma_1(u,x)$ and
\begin{equation}
	\label{MU} \frac{\mathrm{d}m^2 }{\mathrm{d}u} = \frac{2}{X s \nu}\left( sm^2 X\Delta \nu - 4 \alpha \nu \frac{\partial \gamma_1}{\partial x} \right). 
\end{equation}
Making use of the latter equation, we can write $[\delta, \Delta ]\nu$ as
\begin{equation}
	\label{nux} \frac{\partial \Delta \nu/\nu}{\partial x} = - 16\gamma_1 \alpha X 
\end{equation}
and
\begin{equation}
	\label{Deltanunu} \frac{\partial \Delta \nu /\nu}{\partial \nu} = - \frac{\partial \gamma_1 / \partial x}{4 \alpha \nu^3 X }.
\end{equation}
\pagebreak

We now distinguish two subclasses $m = 0$ or $m \neq 0$ (in both classes $b$ and $n$ are arbitrary real functions of $u$) :

\subsubsection{The class $m(u) = 0$}$ $\\
Notice that, if $m(u) = 0$, (\ref{alphaopl}) shows that $\Lambda <0$ and we put $\alpha_0 = \alpha = \sqrt{-2\Lambda}/4$. (\ref{MU}) shows that $\gamma_1$ is independent of $x$. (\ref{Bnu}), (\ref{Cu}) and (\ref{Deltanunu}) then yield
\begin{equation}
	\Delta \nu (u, \nu, x) = - \frac{\nu \partial X/ \partial u}{X}.
\end{equation}
and 
\begin{equation}
	U(u,\nu,x) = \frac{\gamma_1 }{2\alpha_0 \nu} + \nu U_1,
\end{equation}
where $U_1$ is a function of $(u,x)$.

As $\w{\omega}^1$ is now given by 
\begin{equation}
	\w{\omega}^1 = \frac{1}{2} \left(\frac{\mathrm{i}\gamma_1 + \Delta \nu}{2\alpha_0 \nu} +\mathrm{i} \nu U_1 \right) \mathrm{d} u - \frac{1}{4\alpha_0 \nu} \mathrm{d} \nu + 2\mathrm{i} \nu X  \mathrm{d}x,
\end{equation}
with $\gamma_1$ a function of $u$ and $U_1, X$ functions of $(u,x)$, we see that $\mathrm{i} \nu U_1(u,x)$ can be absorbed by a transformation of $x$. The remaining equation to be satisfied (\ref{nux}) then reads
\begin{equation}
	\label{Nx}
	\frac{\partial^2 X}{\partial x \partial u} X - \frac{\partial X}{\partial u} \frac{\partial X}{\partial x} - 16 \gamma_1 \alpha_0 X^3 = 0,
\end{equation}
where there is still freedom in the choice of $x$.
\\

If $\gamma_1 = 0$ the solution of (\ref{Nx}) is of the form $X(u,x) = f(x)/b(u)$, where we can put $f(x)$ equal to 1 by a transformation of $x$. After a transformation of the $r$, $\nu$ and $u$ coordinates, the resulting metric turns out to be precisely the subcase $\gamma_1 = 0$ of (\ref{ds2}) below.\\

If $\gamma_1 \neq 0$, we can use $\Delta \nu/\nu$ as coordinate $x$, (as $[\delta, \Delta] \nu$ and $[\overline{\delta} , \Delta] \nu$ show that $(\delta-\overline{\delta})\Delta\nu\neq 0$, $\nu$, $u$ and $\Delta \nu$ are certainly functionally independent).
From (\ref{nux}) we get $X(u,x)=-1/(16\gamma_1\alpha_0)$, while (\ref{Cu}) and (\ref{Bnu}) yield
\begin{equation}
	U(u,\nu,x) = \frac{\nu}{8\alpha_0 \gamma_1^2} \left(2x\frac{\mathrm{d}\gamma_1}{\mathrm{d}u}-x^2 \gamma_1 +256 b \alpha_0^2 \gamma_1^3  \right)+ \frac{\gamma_1}{2\alpha_0 \nu}.
\end{equation}
Using a coordinate transformation $x\longrightarrow  16 x \gamma_1(u)\alpha_0$, and rescaling $\gamma_1$ and $\nu$, we obtain
\begin{eqnarray}
	\label{ds2}\nonumber
	\mathrm{d}s^2 = \left[ 2n-\Lambda r^2 - \nu + \gamma_1^2\left(\Lambda^3 (b+x^2)^2\nu^2 - 2 \Lambda (b - x^2) +  \frac{1}{\Lambda \nu^2}\right)\right] \mathrm{d} u^2\\ + 2\left[ \mathrm{d} r
	+\frac{4x\gamma_1}{\nu} \mathrm{d} \nu  -  2\gamma_1 \left(\Lambda^2(b+x^2)\nu^2+1 \right) \mathrm{d} x \right] \mathrm{d} u + \Lambda \nu^2 \mathrm{d} x^2+ \frac{\mathrm{d} \nu^2}{ \Lambda \nu^2}. 
\end{eqnarray}

\subsubsection{The class $m(u) \neq 0$}$ $\\
If $m(u)\neq 0$, 
we get, from (\ref{MU}) and (\ref{Deltanunu}), that 
\begin{equation}
	\Delta \nu = \frac{1}{s m^2 X}\left( 4\alpha \nu \frac{\partial \gamma_1}{\partial x} + s m \nu X \frac{\mathrm{d} m}{\mathrm{d} u} \right).
\end{equation} 
(\ref{Bnu}) and (\ref{Cu}) then yield
\begin{equation}
	U = \frac{16 \gamma_1 \alpha \nu }{s m^2},
\end{equation}
\begin{equation}
	X = \frac{X_1(x)}{m(u)},
\end{equation}
and (\ref{nux}) can be rewritten as 
\begin{equation}
	X_1 \frac{\partial^2 \gamma_1}{\partial x^2} - \frac{\partial \gamma_1}{\partial x} \frac{\mathrm{d} X_1}{\mathrm{d} x} +s 4X_1^3 \gamma_1 = 0.
\end{equation}
Applying the coordinate transformation $x \rightarrow y(x)$, with $\frac{\mathrm{d} y }{\mathrm{d} x} = X_1$, the latter equation transforms to
\begin{equation}
	\label{bla}
	\left(\frac{\mathrm{d} y}{\mathrm{d} x}\right) \left( \frac{\partial^2\gamma_1}{\partial y^2} +s 4 \gamma_1\right) = 0,
\end{equation}
which gives 
\begin{eqnarray}
	\gamma_1 &=& P_1(u)\Sigma (2y+P_2(u)),
\end{eqnarray}
with $\Sigma = \sin$ or $\sinh$ according to the sign of $s$. We can now use the freedom in the choice of $r$ to put $P_1(u)$ equal to 1.

After rescaling $\gamma_1$ and applying a transformation $\nu \rightarrow m \nu/\sqrt{2}$, we get the following expression for the line-element :
\begin{eqnarray}
	\label{ds3}
	\nonumber \mathrm{d}s^2&=&\left[ 2n-\Lambda r^2 -4(s- \Lambda \nu^2)\gamma_1^2 + \frac{2 m\sqrt{s-\Lambda \nu^2} }{\Lambda} - \left(\frac{\partial \gamma_1}{\partial y}\right)^2 \right] \mathrm{d}u^2\\
	\nonumber &+&\left[ 2\mathrm{d}r - 4 \nu \sqrt{s-\Lambda \nu^2}\gamma_1 \mathrm{d}y + \frac{2}{\sqrt{s-\Lambda\nu^2} } \frac{\partial \gamma_1}{\partial y} \mathrm{d} \nu\right] \mathrm{d} u \\
	&-& \nu^2 \mathrm{d}y^2 - \frac{1}{s-\Lambda \nu^2}\mathrm{d}\nu^2.
\end{eqnarray}

\subsection{The family $\pi \neq 0$}\label{subsection: pi nonzero}

In this case (\ref{ggg}) yields
\begin{equation*}
	\Phi_{22} = -\frac{\nu(\Psi_2 + \Lambda /3)}{\pi}.
\end{equation*}
The Bianchi equations can be written as follows : \\
\begin{minipage}{0.4\textwidth}
	\begin{eqnarray*}
		\delta \Psi_2 = -3\pi \Psi_2,\\
		\overline{\delta}\Psi_2 = -3\pi \Psi_2,\\
		\Delta \Psi_2 = 0, \\
		\mathrm{D}\Psi_2 = 0,\\
	\end{eqnarray*}
\end{minipage}
\begin{minipage}{0.59\textwidth}
	\begin{eqnarray*}
		\delta \pi = \frac{\pi (\delta \nu - \nu\pi)}{\nu},\\
		\overline{\delta} \pi = \frac{\pi (\overline{\delta} \nu - \nu\pi )}{\nu},\\
		\mathrm{D} \pi = \frac{\pi \mathrm{D} \nu }{\nu},
	\end{eqnarray*}
\end{minipage}
while the NP Ricci equations yield
\begin{equation*}
	\alpha = \frac{3\Psi_2 + \Lambda - 6\pi^2}{12\pi},\qquad \qquad \qquad \overline{\gamma} = \gamma
\end{equation*}
and\\
\begin{minipage}{0.4\textwidth}
	\begin{eqnarray*}
		\delta \nu = -\frac{\nu (3\Psi_2 + \Lambda)}{6\pi},\\
		\overline{\delta}\nu = -\frac{\nu (3\Psi_2 + \Lambda )}{6\pi},\\
		\mathrm{D}\nu = 0,\\
	\end{eqnarray*}
\end{minipage}
\begin{minipage}{0.59\textwidth}
	\begin{eqnarray*}
		\Delta \pi = 0 ,\\
		\delta \gamma = 0,\\
		\overline{\delta}\gamma = 0,\\
		\mathrm{D}\gamma = -c_1\Psi_2^{2/3}.\\
	\end{eqnarray*}
\end{minipage}
We thus have the total derivatives :
\begin{eqnarray*}
	\mathrm{d} \gamma &= \Delta \gamma \w{\omega}^3 -c_1\Psi_2^{2/3} \w{\omega}^4, \\
    \mathrm{d} \nu &= - \frac{\nu(3\Psi_2 + \Lambda)}{6 \pi} (\w{\omega}^1 + \w{\omega}^2) + \Delta \nu \w{\omega}^3, \\
    \mathrm{d} \pi &= -(\pi^2 + \frac{\Psi_2}{2} + \frac{\Lambda}{6}) (\w{\omega}^1 + \w{\omega}^2), \\
    \mathrm{d} \Psi_2 &= -3\pi \Psi_2 (\w{\omega}^1 + \w{\omega}^2).
\end{eqnarray*}
Remember $\Psi_2 = \Psi_2(\pi)$ (\ref{vgl}). We can use $\pi$ (or $\Psi_2$) as a coordinate, but we prefer to write $\pi = \pi(x)$, $\Psi_2 = \Psi_2(x)$ etc.
$ $\\
The Cartan equations and the directional derivatives of $\pi$ show that we can write the basis one-forms as :
\begin{eqnarray}
    \w{\omega}^1 = S\mathrm{d}x + \mathrm{i} (U\mathrm{d}u + X\mathrm{d}x + f \mathrm{d}y),\\
    \w{\omega}^3 = \mathrm{d} u,\\
    \w{\omega}^4 = \mathrm{d}r+ H \mathrm{d} u + F \mathrm{d}x+ G\mathrm{d}y,
\end{eqnarray}
where $f, F, G, H, U$ and $X$ are real functions of $(u,r ,x,y)$, and $S$ is a real function of $x$, which has to satisfy
\begin{equation}
	\label{psix} \frac{\mathrm{d}\Psi_2}{\mathrm{d}x} = -6 S \pi \Psi_2,
\end{equation}
or, equivalently,
\begin{equation}
	\label{pix} \frac{\mathrm{d} \pi}{\mathrm{d}x} = -\frac{1}{3} S ( 6\pi^2 + 3 \Psi_2 +\Lambda ).
\end{equation}
The corresponding directional derivatives are given by
\begin{eqnarray}
	\delta = \frac{1}{2Sf}\left( (XG-fF +\mathrm{i}SG )\frac{\partial }{\partial r} + f\frac{\partial }{\partial x} - (X+\mathrm{i}S)\frac{\partial }{\partial y}\right),\\
	\Delta = \frac{1}{f} \left(f\frac{\partial }{\partial u} + (UG-fH) \frac{\partial }{\partial r} - U \frac{\partial }{\partial y}\right),\\
	\mathrm{D} =\frac{\partial }{\partial r}.
\end{eqnarray}
\\
Applying $[\Delta, \mathrm{D}]$ and $[\delta, \mathrm{D}]$ to $y$ shows $f$, $U$ and $X$ are independent of $r$, so a transformation of $y$ exists making $U=0$. From $[\delta, \Delta]y$ it then follows that $f$ and $X$ are independent of $u$ so a further transformation of $y$ exists, making $X=0$. Next $[\delta, \overline{\delta}]y$, (\ref{vgl}) and (\ref{psix}), show that 
\begin{equation}
	f= \frac{\pi}{\Psi_2^{1/3}}, 
\end{equation}
a factor $f_0 f'(y)$ in $f$ being absorbed in $\mathrm{d}y$.
From the directional derivatives of $\gamma$, we get $\gamma = -c_1\Psi_2^{2/3} r + \gamma'(u,x,y)$, $F(u,r,x,y)=-4\pi(x) rS(x)+F_1(u,x,y)$ and $G(u,r,x,y)=G_1(u,x,y)$, where $\gamma'$ has to satisfy
\begin{equation}
	\label{gammax}
	\frac{\partial \gamma'}{\partial x} = -c_1 \Psi_2^{2/3}F_1, 
\end{equation}
\begin{equation}
	\label{gammay}
	\frac{\partial \gamma'}{\partial y} = -c_1 \Psi_2^{2/3} G_1.
\end{equation}
Notice also that, in case $c_1=0$, $\gamma'$ is a function of $u$ only, such that we can use the remaining boost freedom to put $\gamma'$ equal to zero. This allows us to write $\gamma' = c_1^2 b$.
 
The directional derivatives of $\nu$, together with (\ref{psix}), lead to an expression for $\nu$ : 
\begin{equation}
	\nu =\frac{n(u)}{\pi(x)\Psi_2^{1/3}(x)},
\end{equation}
and from $[\Delta, \mathrm{D}]r$, $[\delta, \Delta]r$ and $[\delta, \overline{\delta}]r$ we find an expression for $H$ : 
\begin{eqnarray}
	H &=& \frac{1}{\Psi_2^{2/3}} \left[ c_1 \left(  - r^2\Psi_2^{4/3} + 2 rc_1 b\Psi_2^{2/3} -c_1^2 b^2 + \frac{\partial b}{\partial u} \right)+ \left( m-n\Psi_2^{1/2} \right)\right],
\end{eqnarray}
where $m(u)$, $n(u)$ and $b(u,x,y)$ are real functions.

If we use $\Psi_2$ as coordinate $x$, and perform a translation of $r \rightarrow r+ c_1 b(u,x,y)/x^{2/3}$, the line-element for this family is given by
\begin{eqnarray}
	\label{metricpic1}
	\nonumber \mathrm{d}s^2&=& -\frac{2}{x^{2/3}} \left(c_1x^{4/3}r^2 + nx^{1/3}-m \right) \mathrm{d}u^2+2\mathrm{d}r \mathrm{d}u + \frac{4r}{3x}\mathrm{d}x\mathrm{d}u \\
	&-& \frac{1}{3x^2\left(6x-\Lambda+6c_1x^{2/3} \right)} \mathrm{d}x^2 - \frac{1}{3x^{2/3}}\left(6x-\Lambda+6c_1x^{2/3} \right) \mathrm{d}y^2.
\end{eqnarray} 
Notice that, in the case where $c_1 \neq 0$, we can fix the boost by requiring $n=1$.

\section{Conclusions}

We can state that all Petrov type $D$ pure radiation metrics, with or without cosmological constant, which admit a non-diverging null congruence, can be written in one of the three forms (\ref{ds2}), (\ref{ds3}) or (\ref{metricpic1}). Metric (\ref{metricpic1}) reduces to the metric of Wils and Van den Bergh\cite{WilsVdB} for $n(u)=0$. One can readily show that the subcase $\gamma_1 =0$ of (\ref{ds2}) is a singular limit of (\ref{ds3}). But whether or not (\ref{ds2}) with $\gamma_1 \neq 0$ is a singular limit of (\ref{ds3}) is not yet clear. 
It is clear that in general (\ref{metricpic1}) admits at least one Killing vector field. More isometries may arise for particular choices of the parameters and free functions. A detailed discussion of the physical properties of these metrics will be given in a follow up paper.


\section*{References}


\providecommand{\newblock}{}

\end{document}